\documentclass[b5paper,twoside,reqno]{bjp}
\usepackage{cite}
\usepackage{latexsym}
\usepackage[pdftex]{graphicx}
\usepackage{amssymb,amsmath,amsfonts,amsbsy, amsthm, xspace, latexsym, amscd, enumitem}
\usepackage[utf8]{inputenc}
\usepackage[T1]{fontenc}
\usepackage[english]{babel}
\usepackage{times}
\usepackage{url,bm}
\newcommand\rf[1]{(\ref{eq:#1})}
\newcommand\lab[1]{\label{eq:#1}}
\newcommand\nonu{\nonumber}
\newcommand\br{\begin{eqnarray}}
\newcommand\er{\end{eqnarray}}
\newcommand\be{\begin{equation}}
\newcommand\ee{\end{equation}}

\newcommand\foot[1]{\footnotemark\footnotetext{#1}}
\newcommand\lb{\lbrack}
\newcommand\rb{\rbrack}

\newcommand\llb{\left\lbrack}
\newcommand\rrb{\right\rbrack}

\renewcommand\({\left(}
\renewcommand\){\right)}
\newcommand\bv{\bigm\vert}               

\newcommand\bc{\begin{center}}
\newcommand\ec{\end{center}}

\newcommand\partder[2]{\frac{{\partial {#1}}}{{\partial {#2}}}}

\renewcommand\d{\delta}

\newcommand\eps{\epsilon}

\newcommand\g{\gamma}

\newcommand\h{\frac{1}{2}}

\renewcommand\l{\lambda}

\newcommand\m{\mu}
\newcommand\n{\nu}

\newcommand\vp{\varphi}

\newcommand\pa{\partial}

\newcommand\s{\sigma}

\renewcommand\t{\tau}
\renewcommand\th{\theta}

\newcommand\cJ{{\mathcal J}}

\newcommand\cV{{\mathcal V}}

\newcommand{\ct}[1]{\cite{#1}}
\newcommand{\bib}[1]{\bibitem{#1}}

\newcommand\xdot{\stackrel{.}{x}}

\renewcommand\congress{\small\sf Section: Mathematical Physics}

\makeindex

\begin{document}

\sloppy \raggedbottom

\title{Einstein-Rosen ``Bridge'' Revisited and Lightlike Thin-Shell Wormholes}

\runningheads{Einstein-Rosen ``Bridge'' Revisited}{E. Guendelman, 
E. Nissimov, S. Pacheva, M. Stoilov}

\begin{start}
\coauthor{Eduardo Guendelman}{1}, \coauthor{Emil Nissimov}{2}, 
\coauthor{Svetlana Pacheva}{2}, \author{Michail Stoilov}{2}

\address{Department of Physics, Ben-Gurion University of the Negev, \\
Beer-Sheva 84105, Israel}{1}

\address{Institute for Nuclear Research and Nuclear Energy,\\ 
Bulgarian Academy of Sciences, Sofia 1784, Bulgaria}{2}

\begin{Abstract}
We study in some detail the properties of the mathematically correct
formulation of the classical Einstein-Rosen ``bridge'' as proposed in the
original 1935 paper, which was shown in a series of previous papers of ours
to represent the simplest example of a static spherically symmetric traversable
lightlike thin-shell wormhole. Thus, the original Einstein-Rosen ``bridge''
is {\em not equivalent} to the concept of the {\em dynamical} and 
{\em non-traversable} Schwarzschild wormhole, also called ``Einstein-Rosen bridge''
in modern textbooks on general relativity. The original Einstein-Rosen
``bridge'' requires the presence of a special kind of ``exotic'' matter source 
located on its throat which was shown to be the simplest member of the
previously introduced by us class of lightlike membranes. We introduce and
exploit the Kruskal-Penrose description of the original Einstein-Rosen
``bridge''. In particular, we explicitly construct closed timelike geodesics
on the pertinent Kruskal-Penrose manifold. 
\end{Abstract}


\end{start}

\section{Introduction}
\label{intro}

The celebrated Einstein-Rosen ``bridge'' in its original formulation from 1935 
\ct{ER-1935} is historically the first example of a traversable gravitational
wormhole spacetime. However, the traditional presentation of the Einstein-Rosen 
``bridge'' in modern textbooks in general relativity (\textsl{e.g.} \ct{MTW})
{\em does not} correspond to its original formulation \ct{ER-1935}. 
The ``textbook'' version of the Einstein-Rosen 
``bridge'' is physically inequivalent to the original 1935 construction as 
it represents both a non-static spacetime geometry as well as it is 
{\em non-traversable}. 

Based on earlier works of ours \ct{our-ER-bridge} we revisit the original 
Einstein-Rosen 
formulation from 1935 \ct{ER-1935}. We find that the originally used by Einstein
and Rosen local spacetime coordinates suffer from a serious problem --
the pertinent spacetime metric in these coordinates possesses an essential
unphysical singularity at the wormhole ``throat'' -- the boundary between 
the two ``universes'' of the Einstein-Rosen ``bridge'' manifold.

We proposed instead a different set of local coordinates for the Einstein-Rosen 
``bridge'' such that its spacetime geometry becomes well-defined everywhere, 
including on the wormhole "throat". 

On the other hand, this reveals a very important {\em new feature of 
the correctly defined} Einstein-Rosen ``bridge'' \ct{our-ER-bridge}, which was 
{\em overlooked in the original} Einstein-Rosen paper \ct{ER-1935}. 
Namely, we show that the correct construction of the 
Einstein-Rosen ``bridge'' as self-consistent solution of the corresponding 
Einstein equations requires the presence of a ``thin-shell'' ``exotic'' matter 
source on the wormhole ``throat'' -- a special particular member of the 
originally introduced in other papers of ours \ct{our-LLB,our-ER-bridge}
class of {\em lightlike membranes} \foot{For a detailed discussion of {\em
timelike} thin-shell wormholes, see the book \ct{visser-book}.}.

In the present note we first briefly review the basics of our construction
of the original Einstein-Rosen ``bridge'' as a specific well-defined solution 
of wormhole type of gravity interacting self-consistently with a dynamical
lightlike membrane matter based on explicit Lagrangian action principle for
the latter \ct{our-LLB,our-ER-bridge}. Also we present the maximal analytic
Kruskal-Penrose extension of the original Einstein-Rosen ``bridge'' wormhole 
manifold significantly different from the Kruskal-Penrose manifold of the 
corresponding Schwarzschild black hole \ct{MTW}.

Next, we discuss in some detail the dynamics of test particles (massless and 
massive ones) in the gravitational background of Einstein-Rosen ``bridge'' 
wormhole. Apart from exhibiting the traversability of the Einstein-Rosen 
wormhole w.r.t. proper-time of travelling observers, we explicitly construct
a closed timelike geodesics.

\section{Deficiency of the Original 1935 Formulation of Einstein-Rosen 
Bridge}
\label{deficiency}

The Schwarzschild spacetime metric -- the simplest static spherically symmetric 
black hole metric --  is given in standard coordinates $(t,r,\th,\vp)$ as
(\textsl{e.g.} \ct{MTW}):
\be
ds^2 = - A(r) dt^2 + \frac{1}{A(r)} dr^2 + r^2 \( d\th^2 + \sin^2\th d\vp^2\) 
\quad ,\quad A(r) = 1 - \frac{r_0}{r} \; .
\lab{Schw-metric}
\ee
$r_0 \equiv 2m$ ($m$ -- black hole mass parameter) is the horizon
radius, where $A(r_0) = 0$ ($r=r_0$ is a non-physical coordinate singularity of the metric \rf{Schw-metric}, unlike
the physical spacetime singularity at $r=0$). Here 
$r>r_0$ defines the exterior Schwarzschild spacetime region, 
whereas the region $r<r_0$ is the black hole interior.

In spacetime geometries of static spherically symmetric type (like
\rf{Schw-metric} with generic $A(r)$) special role is being played by the
so called ``tortoise'' coordinate $r^{*}$ defined as:
\be
\frac{dr^{*}}{dr}=\frac{1}{A(r)} \quad \longrightarrow \quad
r^{*} = r + r_0 \log\bigl(|r-r_0|/r_0\bigr) \; ,
\lab{tortoise}
\ee
such that for radially moving light rays we have $t\pm r^{*}={\rm const}$
(curved spacetime generalization of Minkowski's $t\pm r={\rm const}$).

In constructing the maximal analytic extension of the Schwarzschild spacetime 
geometry -- the {\em Kruskal-Szekeres} coordinate chart -- essential intermediate
use is made of ``tortoise'' coordinate $r^{*}$ \rf{tortoise}, where
the Kruskal-Szekeres (``light-cone'') coordinates $(v,w)$ 
are defined as follows (\textsl{e.g.} \ct{MTW}):
\be
v = \pm \frac{1}{\sqrt{2k_h}} e^{k_h \bigl(t+r^{*}\bigr)} \quad ,\quad
w = \mp \frac{1}{\sqrt{2k_h}} e^{-k_h \bigl(t-r^{*}\bigr)} 
\lab{KS-coord}
\ee
with all four combinations of the overall signs. Here
$k_h = \h \pa_r A(r)\bv_{r=r_0} = \frac{1}{2r_0}$ denotes the so called
``surface gravity'', which is related to the Hawking temperature as
$\frac{k_h}{2\pi}=k_B T_{\rm hawking}$. Equivalently, Eqs.\rf{KS-coord} can
be writtes as::
\be
\mp vw = \frac{1}{k_h} e^{2 k_h r^{*}} \quad ,\quad \mp\frac{v}{w} = e^{2 k_h t} \; ,
\lab{KS-coord-1}
\ee
wherefrom $t$ and $r^{*}$, as well as $r$, are determined as functions of $vw$. 

The various combination of the overall signs in Eqs.\rf{KS-coord} define a
doubling of the two regions of the standard Schwarzschild geometry \ct{MTW}:

(i) $(+,-)$ -- exterior Schwarzschild region $r>r_0$ (region $I$); 

(ii) $(+,+)$ -- black hole $r<r_0$ (region $II$); 

(iii) $(-,+)$ -- second copy of exterior Schwarzschild region $r>r_0$ (region $III$); 

(iv) $(-,-)$ -- ``white'' hole region $r<r_0$ (region $IV$).


In the classic paper \ct{ER-1935} Einstein and Rosen introduced in \rf{Schw-metric}
a new radial-like coordinate $u$ via $r = r_0 + u^2$:
\be
ds^2 = - \frac{u^2}{u^2 + r_0} dt^2 + 4 (u^2 + r_0)du^2 +
(u^2 + r_0)^2 \( d\th^2 + \sin^2 \th \,d\vp^2\) \; ,
\lab{E-R-metric}
\ee
and let $u \in (-\infty, +\infty)$. Therefore, \rf{E-R-metric} describes two 
identical copies of the exterior Schwarzschild spacetime region ($r> r_0$) 
for $u>0$ and $u<0$, respectively, which are formally glued together at the 
horizon $u=0$.

However, there is a very serious problem with \rf{E-R-metric} (apart from the
coordinate singularity at $u=0$, where $\det\Vert g_{\m\n}\Vert_{u=0} = 0$).
The Einstein-Rosen metric \rf{E-R-metric} {\em does not} satisfy the vacuum
Einstein equations at $u=0$. The latter acquire an {\em ill-defined} non-vanishing 
``matter'' stress-energy tensor term on the r.h.s., which was overlooked in 
the original 1935 paper \ct{ER-1935}!

Indeed, as explained in \ct{our-ER-bridge}, using Levi-Civita identity 
$R^0_0 = - \frac{1}{\sqrt{-g_{00}}} \nabla^2_{(3)} \(\sqrt{-g_{00}}\)$
(where $\nabla^2_{(3)}$ is the 3-dimensional spatial Laplacian) we
deduce that \rf{E-R-metric} solves vacuum Einstein equation $R^0_0 = 0$ for all
$u\neq 0$. However, since $\sqrt{-g_{00}} \sim |u|$ as $u \to 0$ and since
$\frac{\pa^2}{{\pa u}^2} |u| = 2 \d (u)$, Levi-Civita identity tells us that:
\be
R^0_0 \sim \frac{1}{|u|} \d (u) \sim \d (u^2) \; ,
\lab{ricci-delta}
\ee
and similarly for the scalar curvature $R \sim \frac{1}{|u|} \d (u) \sim \d (u^2)$.

\section{Original Einstein-Rosen Bridge is a Lightlike Thin-Shell Wormhole}
\label{ER-bridge+LL-brane}

In Refs.\ct{our-ER-bridge} we proposed a correct reformulation of the original
Einstein-Rosen bridge as a mathematically consistent traversable 
lightlike thin-shell wormhole. This is achieved via introducing a different
radial-like coordinate $\eta \in (-\infty, +\infty)$, by
substituting $r = r_0 + |\eta|$ in \rf{Schw-metric}:
\be
ds^2 = - \frac{|\eta|}{|\eta| + r_0} dt^2 + \frac{|\eta| + r_0}{|\eta|}d\eta^2 +
(|\eta| + r_0)^2 \( d\th^2 + \sin^2 \th \,d\vp^2\) \; .
\lab{our-ER-metric}
\ee
Obviously, Eq.\rf{our-ER-metric} again describes two ``universes'' -- 
two identical copies of the exterior Schwarzschild spacetime region for 
$\eta >0$ and $\eta <0$, respectively. However, unlike the ill-behaved
original 1935 metric \rf{E-R-metric}, now both ``universes'' are correctly 
glued together at their common horizon $\eta=0$.

Namely, the metric \rf{our-ER-metric} solves Einstein equations:
\be
R_{\m\n} - \h g_{\m\n} R = 8\pi T^{(brane)}_{\m\n} \; ,
\lab{Einstein-eqs+LL}
\ee
where on the r.h.s. $T^{(brane)}_{\m\n} = S_{\m\n} \d (\eta)$ is
the energy-momentum tensor of a special kind of a {\em lightlike membrane}
located on the common horizon $\eta=0$ -- the wormhole ``throat''.
As shown in \ct{our-ER-bridge}, the lightlike analogues of W.Israel's 
junction conditions on the 
wormhole ``throat'' are satisfied. The resulting lightlike thin-shell wormhole 
is traversable (see Sections 4,6 below).

The energy-momentum tensor of lightlike membranes $T^{(brane)}_{\m\n}$ is 
self-consistently derived as 
$T^{(brane)}_{\m\n} = - \frac{2}{\sqrt{-g}} \frac{\d S_{\rm LL}}{\d g^{\m\n}}$
from the following manifestly reparametrization invariant world-volume 
Polyakov-type lightlike membrane action (written for arbitrary $D=(p+1)+1$ 
embedding spacetime dimension and $(p+1)$-dimensional brane world-volume):
\br
S_{\rm LL} = - \h \int d^{p+1} \s\, T b_0^{\frac{p-1}{2}}\sqrt{-\g}
\llb \g^{ab} {\bar g}_{ab} - b_0 (p-1)\rrb \; ,
\lab{LL-action} \\
{\bar g}_{ab} \equiv 
g_{ab} - \frac{1}{T^2} \pa_a u \pa_b u \quad ,\quad 
g_{ab} \equiv \pa_a X^{\m} g_{\m\n}(X) \pa_b X^{\n} \; .
\lab{ind-metric-ext}
\er
Here the following notations are used:

${}\quad{}$ (a) $\g_{ab}$ is the {\em intrinsic} Riemannian metric on the world-volume with
$\g = \det \Vert\g_{ab}\Vert$;
$b_0$ is a positive constant measuring the world-volume ``cosmological constant''.
$g_{ab}$ \rf{ind-metric-ext} is the {\em induced} metric on the world-volume 
which becomes {\em singular} on-shell --  manifestation of the lightlike 
nature of the $p$-brane.

${}\quad{}$ (b) $(\s)\equiv \(\s^a \)$ with $a=0,1,\ldots ,p$ ; $\pa_a \equiv \partder{}{\s^a}$.

${}\quad{}$ (c) $X^\m (\s)$ are the $p$-brane embedding coordinates in the bulk
$D$-dimensional spacetime with Riemannian metric 
$g_{\m\n}(x)$ ($\m,\n = 0,1,\ldots ,D-1$). 

${}\quad{}$ (d) $u$ is auxiliary world-volume scalar field defining the lightlike direction
of the induced metric $g_{ab}$ \rf{ind-metric-ext} 
and it is a non-propagating degree of freedom. 

${}\quad{}$ (e) $T$ is {\em dynamical (variable)} membrane tension 
(also a non-propagating degree of freedom).


The Einstein Eqs.\rf{Einstein-eqs+LL} imply the following relation between 
the lightlike membrane parameters and the Einstein-Rosen bridge ``mass'' 
($r_0 = 2m$):
\be
-T=\frac{1}{8\pi m} \;\; ,\;\; b_0 = \frac{1}{4} \; ,
\lab{rel-param}
\ee
\textsl{i.e.}, the lightlike membrane dynamical tension $T$ becomes {\em negative}
on-shell -- manifestation of the ``exotic matter'' nature of the lightlike
membrane.


\section{Test Particle Dynamics and Traversability in the Original 
Einstein-Rosen Bridge}
\label{test-particle}

As already noted in \ct{our-ER-bridge} traversability of the
original Einstein-Rosen bridge is a particular manifestation of the
traversability of lightlike ``thin-shell'' wormholes \foot{Subsequently,
traversability of the Einstein-Rosen bridge has been studied using
Kruskal-Szekeres coordinates for the Schwarzschild black hole \ct{poplawski}, 
or the 1935 Einstein-Rosen coordinate chart \rf{E-R-metric} \ct{katanaev}.}.
Here for completeness we will present the explicit details of the
traversability within the proper Einstein-Rosen bridge wormhole coordinate
chart \rf{our-ER-metric} which are needed for the construction of the
pertinent Kruskal-Penrose diagram in Section 4.

The motion of test-particle (``observer'') of mass $m_0$ in a gravitational background 
is given by the reparametrization-invariant world-line action:
\be
S_{\rm particle} = 
\h \int d\l \llb \frac{1}{e} g_{\m\n} \xdot^\m \xdot^\n - e m_0^2\rrb \; ,
\lab{particle-action}
\ee
where $\xdot^\m \equiv \frac{dx^\m}{d\l}$, $e$ is the world-line ``einbein'' and
in the present case $(x^\m) = (t,\eta,\th,\vp)$.

For a static spherically symmetric background such as \rf{our-ER-metric} 
there are conserved Noether ``charges'' -- energy $E$ and angular momentum $\cJ$. 
In what follows we will consider purely ``radial'' motion ($\cJ=0$) so, upon taking 
into account the ``mass-shell'' constraint (the equation of motion w.r.t. $e$) 
and introducing the world-line proper-time parameter $\t$ ($\frac{d\t}{d\l}=e m_0$), 
the timelike geodesic equations (world-lines of massive point particles) read: 
\be
\Bigl(\frac{d\eta}{d\t}\Bigr)^2 = \frac{E^2}{m_0^2} - A(\eta) \;\; ,\;\;
\frac{dt}{d\t}= \frac{E}{m_0 A(\eta)} \;\; ,\;\; 
A(\eta) \equiv \frac{|\eta|}{|\eta| + r_0} \; .
\lab{radial-eqs}
\ee
where $A(\eta)$ is the ``$-g_{00}$'' component of the proper Einstein-Rosen 
bridge metric \rf{our-ER-metric}.


The first radial $\eta$-equation \rf{radial-eqs} exactly resembles classical
energy-conservation equation for a ``non-relativistic'' particle with mass
$\h$ moving in an effective potential $\cV_{\rm eff} (\eta) \equiv\h A(\eta)$ graphically
depicted on Fig.1 below:
\be
\frac{d\eta}{d\t} = \eps \sqrt{\frac{E^2}{m_0^2} - A(\eta)} \quad ,\quad
\eps = \pm 1 \;,
\lab{eta-eq-0}
\ee
depending whether $\eta (\t)$ moves towards larger values ($\eps=+1$) or
lower values ($\eps=-1$).

For a test-particle starting for $\t=0$ at initial position 
$\eta_0 = \eta (0)\; , t_0 = t(0)$
the solutions of Eqs.\rf{radial-eqs} read:
\br
\eps \frac{m_0}{2k_h E} \int^{2k_h \eta_(\t)}_{2k_h \eta_0} dy
\sqrt{(1+|y|)\Bigl\lb(1+\bigl(1-\frac{m_0^2}{E^2}\bigr)|y|\Bigr\rb^{-1}}
= \t \; ,
\lab{eta-eq}\\
\eps \frac{1}{2k_h} \int^{2k_h \eta (\t)}_{2k_h \eta_0} dy \frac{1}{|y|} 
\sqrt{(1+|y|)^3\Bigl\lb(1+\bigl(1-\frac{m_0^2}{E^2}\bigr)|y|\Bigr\rb^{-1}}
= t(\t)-t_0 \; .
\lab{t-eq}
\er
In particular, Eq.\rf{eta-eq} shows that the particle will cross the wormhole
``throat'' ($\eta=0$) within a finite proper-time $\t_0 >0$:
\be
\t_0 = \eps \frac{m_0}{2k_h E} \int^{0}_{2k_h \eta_0} dy
\sqrt{(1+|y|)\Bigl\lb(1+\bigl(1-\frac{m_0^2}{E^2}\bigr)|y|\Bigr\rb^{-1}}
\lab{eta-eq-00}
\ee
(here $\eps = -1$ for $\eta_0 >0$ and $\eps = +1$ for $\eta_0 <0$).

Concerning the ``laboratory'' time $t$, it follows from \rf{t-eq} 
that $t(\t_0 - 0)=+\infty$, \textsl{i.e.}, from the point of view of a 
static observer in ``our'' (right) universe it will take infinite 
``laboratory'' time for the particle to reach the ``throat'' -- the latter 
appears to the static observer as a future black hole horizon.

Eq.\rf{t-eq} also implies $t(\t_0 + 0)= -\infty$, which means that from the 
point of view of a static observer in the second (left) universe, upon 
crossing the ``throat'', the particle starts its motion in the second (left) 
universe from infinite past, so that it will take an infinite amount of 
``laboratory'' time to reach a point $\eta_1 <0$ -- \textsl{i.e.} the 
``throat'' now appears as a past black hole horizon.

For small energies $E < m_0$ according to \rf{eta-eq} the particle is trapped in
an effective potential well and shuttles within finite proper-time intervals
between the ``reflection'' points (see Fig.1):
\be
\pm \eta_{\rm stop}= \Bigl(2 k_H \lb m_0^2/E^2 -1\rb\Bigr)^{-1} \; .
\lab{eta-stop}
\ee
In Section 6 we will show that for a special value of $m_0/E$ 
the pertinent particle geodesics is a closed 
timelike curve on the extended Kruskal-Penrose manifold.

\begin{figure}
\begin{center}
\includegraphics[width=10cm,keepaspectratio=true]{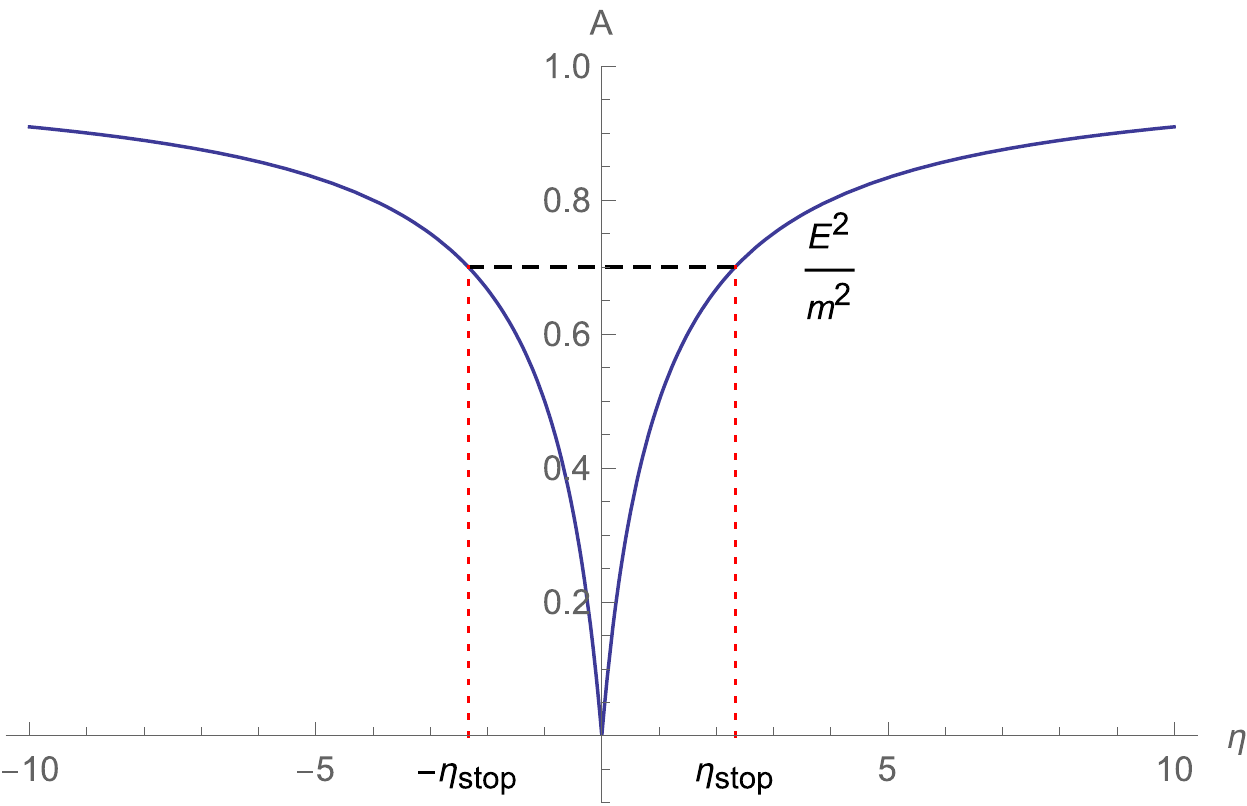}
\caption{Graphical representation of $A(\eta)$ \rf{radial-eqs} with 
``reflection'' points $\pm \eta_{\rm stop}$ \rf{eta-stop} indicated.}
\end{center}
\end{figure}

In analogy with the usual ``tortoise'' coordinate $r^{*}$ for the
Schwarzschild black hole geometry \rf{tortoise} we now introduce
Einstein-Rosen bridge ``tortoise'' coordinate $\eta^{*}$ 
(recall $r_0 = \frac{1}{2k_h}$):
\be
\frac{d\eta^{*}}{d\eta} = \frac{|\eta|+r_0}{|\eta|} \quad \longrightarrow
\quad \eta^{*} = \eta + {\rm sign} (\eta) r_0 \log \bigl(|\eta|/r_0\bigr)
\; .
\lab{ER-tortoise}
\ee
Let us note here an important difference in the behavior of the
``tortoise'' coordinates $r^{*}$ \rf{tortoise} and $\eta^{*}$
\rf{ER-tortoise} in the vicinity of the horizon. Namely:
\be
r^{*} \to -\infty \quad {\rm for} \;\; r \to r_0 \pm 0 \; ,
\lab{tortoise-r}
\ee
\textsl{i.e.}, when $r$ approaches the horizon either from above or from below,
whereas when $\eta$ approaches the horizon from above or from below:
\be
\eta^{*} \to \mp \infty \quad {\rm for} \;\; \eta \to \pm 0 \; .
\lab{tortoise-eta}
\ee

For infalling/outgoing massless particles (light rays) 
Eqs.\rf{eta-eq}-\rf{ER-tortoise} imply:
\be
t\pm \eta^{*} = {\rm const} \; .
\lab{t-plus-eta-star-1}
\ee
For infalling/outgoing massive particles we obtain accordingly:
\be
\bigl\lb t\pm\eta^{*}\bigr\rb (\t) = \frac{1}{2k_h}
\int^{2k_h \eta (\t)}_{2k_h \eta_0} dx \Bigl( 1+ \frac{1}{|x|}\Bigr)
\llb \eps\sqrt{(1+|x|)\Bigl\lb(1+\bigl(1-\frac{m_0^2}{E^2}\bigr)|x|\Bigr\rb^{-1}}
\pm 1 \rrb \; .
\lab{t-plus-eta-star-2}
\ee

\section{Kruskal-Penrose Manifold of the Original Einstein-Rosen Bridge}
\label{kruskal-penrose}

Following \ct{ER-LT-11} we now introduce the maximal analytic extension of original 
Einstein-Rosen wormhole geometry \rf{our-ER-metric} through the following
Kruskal-like coordinates $(v,w)$:
\be
v = \pm \frac{1}{\sqrt{2k_h}} e^{\pm k_h (t+\eta^{*})} \;\; , \;\;
w = \mp \frac{1}{\sqrt{2k_h}} e^{\mp k_h (t-\eta^{*})} \; ,
\lab{region-1-2}
\ee
implying:
\be
-vw = \frac{1}{2k_h} e^{\pm 2k_h \eta^{*}} \;\; ,\;\;
-\frac{v}{w} = e^{\pm 2k_h t} \; .
\lab{region-1-2-a}
\ee
Here and below $\eta^{*}$ is given by \rf{ER-tortoise}.

The upper signs in \rf{region-1-2}-\rf{region-1-2-a} correspond to region $I$
$(v>0,w<0)$ describing ``our'' (right) universe $\eta>0$, whereas the
lower signs in \rf{region-1-2}-\rf{region-1-2-a} correspond to region $II$ 
$(v<0,w>0)$ describing the second (left) universe $\eta<0$ (see Fig.2 below).

Using the explicit expression \rf{ER-tortoise} for $\eta^{*}$ in 
\rf{region-1-2-a} we find two ``throats'' (horizons) -- at $v=0$ or $w=0$
corresponding to $\eta=0$:

${}\quad{}$ (a) In region $I$ the ``throat'' $(v>0,w=0)$ is a future horizon
$(\eta=0\, ,\, t\to +\infty)$, whereas the ``throat'' $(v=0,w<0)$ is a past horizon
$(\eta=0\, ,\, t\to -\infty)$.

${}\quad{}$ (b) In region $II$ the ``throat'' $(v=0,w>0)$ is a future horizon
$(\eta=0\, ,\, t\to +\infty)$, whereas the ``throat'' $(v<0,w=0)$ is a past horizon
$(\eta=0\, ,\, t\to -\infty)$.

As usual one replaces Kruskal-like coordinates $(v,w)$ \rf{region-1-2} with 
compactified Penrose-like coordinates $({\bar v},{\bar w})$:
\br
{\bar v} = \arctan (\sqrt{2k_h}\, v) = 
\pm \arctan\bigl( e^{\pm k_h (t+\eta^{*})}\bigr) \; ,
\nonu \\
{\bar w} = \arctan (\sqrt{2k_h}\, w)= 
\mp \arctan\bigl( e^{\mp k_h (t-\eta^{*})}\bigr) \; ,
\lab{Penrose-coord}
\er
mapping the various ``throats'' (horizons) and infinities to 
finite lines/points:

\begin{itemize}
\item
In region $I$: future horizon $(0<{\bar v}<\frac{\pi}{2}, {\bar w}=0)$; 
past horizon $({\bar v}=0, -\frac{\pi}{2}<{\bar w} <0)$.
\item
In region $II$: future horizon $({\bar v}=0, 0<{\bar w}<\frac{\pi}{2})$; 
past horizon $(-\frac{\pi}{2}<{\bar v}<0, {\bar w}=0)$.
\item
$i_0$ -- spacelike infinity ($t={\rm fixed}, \eta \to \pm\infty$):\\
$i_0 = (\frac{\pi}{2},-\frac{\pi}{2})$ in region $I$;
$i_0 = (-\frac{\pi}{2},\frac{\pi}{2})$ in region $II$.
\item
$i_{\pm}$ -- future/past timelike infinity ($t \to \pm\infty, \eta={\rm fixed}$):\\
$i_{+}=(\frac{\pi}{2},0)$, $\, i_{-}=(0, -\frac{\pi}{2})$ in region $I$;
$i_{+}=(0,\frac{\pi}{2})$, $\, i_{-}=(-\frac{\pi}{2},0)$ in region $II$.
\item
$J_{+}$ -- future lightlike infinity ($t\to +\infty, \eta \to \pm\infty$, 
$\; t \mp \eta^{*} = {\rm fixed}$):\\
$J_{+} = ({\bar v}=\frac{\pi}{2}, -\frac{\pi}{2}<{\bar w}<0)$ in region $I$;\\
$J_{+} = (-\frac{\pi}{2}<{\bar v}<0,{\bar w}=\frac{\pi}{2})$ in region $II$.
\item
$J_{-}$ -- past lightlike infinity ($t\to -\infty, \eta \to \pm\infty$), 
$\; t \pm \eta^{*} = {\rm fixed}$):\\
$J_{-} = (0<{\bar v}<\frac{\pi}{2},{\bar w}= -\frac{\pi}{2})$ in region $I$:\\
$J_{-} = ({\bar v}= -\frac{\pi}{2},0<{\bar w}<\frac{\pi}{2})$ in region $II$.
\end{itemize}

In Ref.\ct{ER-LT-11}, using the continuity of the light ray geodesics 
\rf{t-plus-eta-star-1} when starting in one of the regions $I$ or $II$ and
crossing the horizon (``throat'') into the other one,  we have exhibited
the following mutual identification of a future horizon of one region
with the past horizon of the second region (see Fig.2):

\begin{figure}
\begin{center}
\includegraphics[width=13cm,keepaspectratio=true]{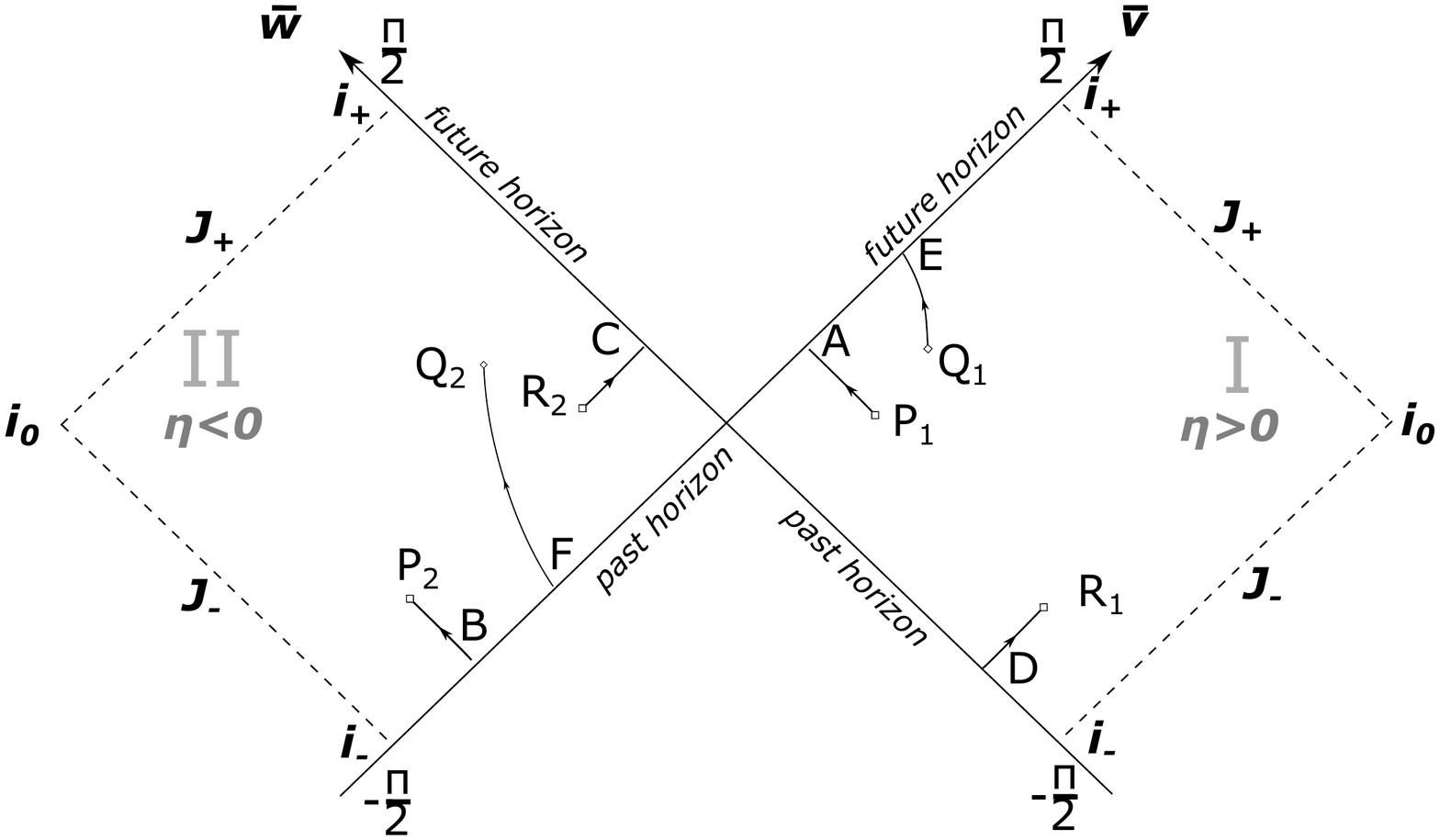}
\caption{Kruskal-Penrose Manifold of the Original Einstein-Rosen Bridge}
\end{center}
\end{figure}

\begin{itemize}
\item
Future horizon in region $I$ is identified with past horizon in region $II$ as:
\be
({\bar v},0) \sim ({\bar v}-\frac{\pi}{2},0)\; .
\lab{ident-1}
\ee
Infalling light rays cross from region $I$ into region $II$ via paths 
$P_1 \to A\sim B \to P_2$ -- all the way within finite world-line time intervals
(the symbol $\sim$ means identification according to \rf{ident-1}).
Similarly, infalling massive particles cross from region $I$ into region $II$ 
via paths $ Q_1 \to E \sim F \to Q_2$ within finite proper-time interval.
\item
Future horizon in $II$ is identified with past horizon in $I$:
\be
(0,{\bar w}) \sim (0, {\bar w}-\frac{\pi}{2}) \; .
\lab{ident-2}
\ee
Infalling light rays cross from region $II$ into
region $I$ via paths $R_2 \to C\sim D \to R_1$ where $C\sim D$ is identified
according to \rf{ident-2}.
\end{itemize}

\section{Closed Timelike Geodesics}
\label{CTC}

Let us consider again massive test-particle dynamics with small 
energies $E < m_0$, which according to \rf{eta-eq} means that the particle 
is trapped within an effective potential well (Fig.1).

Let the particle starts in ``universe'' $I$ at $\eta=0$ (past horizon) (at
some point $A$ of the Kruskal-Penrose coordinate chart as indicated on 
Fig.3) and then moves radially forward in $\eta$ until it reaches the 
``reflection'' point $\eta_{\rm stop}$ \rf{eta-stop} (indicated by $Q_1$ on
Fig.3) within the finite proper-time interval according to \rf{eta-eq}:
\br
\Delta\t = \frac{m_0}{2k_h E} \int^{2k_h \eta_{\rm stop}}_{0} dx
\sqrt{(1+|x|)\Bigl\lb 1+\bigl(1-\frac{m_0^2}{E^2}\bigr)|x|\Bigr\rb^{-1}}
\nonu \\
= \frac{m_0}{2k_h E} \Bigl\lb \frac{1}{b} + 
\frac{b+1}{b^{3/2}}\Bigl(\frac{\pi}{2} - \arctan(\sqrt{b})\Bigr)\Bigr\rb
\; ,
\lab{delta-tau}
\er
where a short hand notation $b$ is introduced:
\be
b \equiv \frac{m^2_0}{E^2} - 1 > 0 \; .
\lab{b-def}
\ee
Then the particle proceeds by returning backward in $\eta$ from 
$\eta_{\rm stop}$ \rf{eta-stop} towards $\eta=0$ (future horizon of
Kruskal-Penrose region $I$) within the same proper-time interval
\rf{delta-tau} and it crosses the horizon at the point $B$ on Fig.3.
Thus, the particle enters Kruskal-Penrose region $II$ ($\eta=0$ -- past
horizon in $II$) at the point $C$ on Fig.3 which is ``dual'' to point $B$ on
the future horizon of $I$ according to the future/past horizon identification
\rf{ident-1}-\rf{ident-2}. Now the particle continues towards negative
$\eta$ until it reaches within the same proper-time interval \rf{delta-tau}
the other ``reflection point'' $-\eta_{\rm stop}$ \rf{eta-stop} (indicated 
by $Q_2$ on Fig.3). Finally, the particle returns from $-\eta_{\rm stop}$ 
\rf{eta-stop} towards $\eta=0$ (the future horizon of $II$) and reaches it at
the point $D$ on the Kruskal-Penrose chart (Fig.3) within the same
proper-time interval \rf{delta-tau}. Then it crosses 
into region $I$ at the point $F$ on the past horizon of $I$, which 
is dually equivalent to $D$ on the future horizon of $II$ 
according to the horizon identification \rf{ident-1}-\rf{ident-2}.
Afterwards the particle continues again from $F$ towards the first 
``reflection'' point $Q_1$ on Fig.3.

We want now to find the conditions for the coincidence of the points
$F = A$, \textsl{i.e.}, to find conditions for the existence of a
closed timelike curve (CTC) meaning that the particle travels from some
starting spacetime point in ``universe'' $I$, crosses into ``universe'' 
$II$, then crosses back into ``universe'' $I$ and returns to 
the {\em same} starting spacetime point for a {\em finite} proper-time
interval equal to $4 \Delta\t$ \rf{delta-tau}.

\begin{figure}
\begin{center}
\includegraphics[width=13cm,keepaspectratio=true]{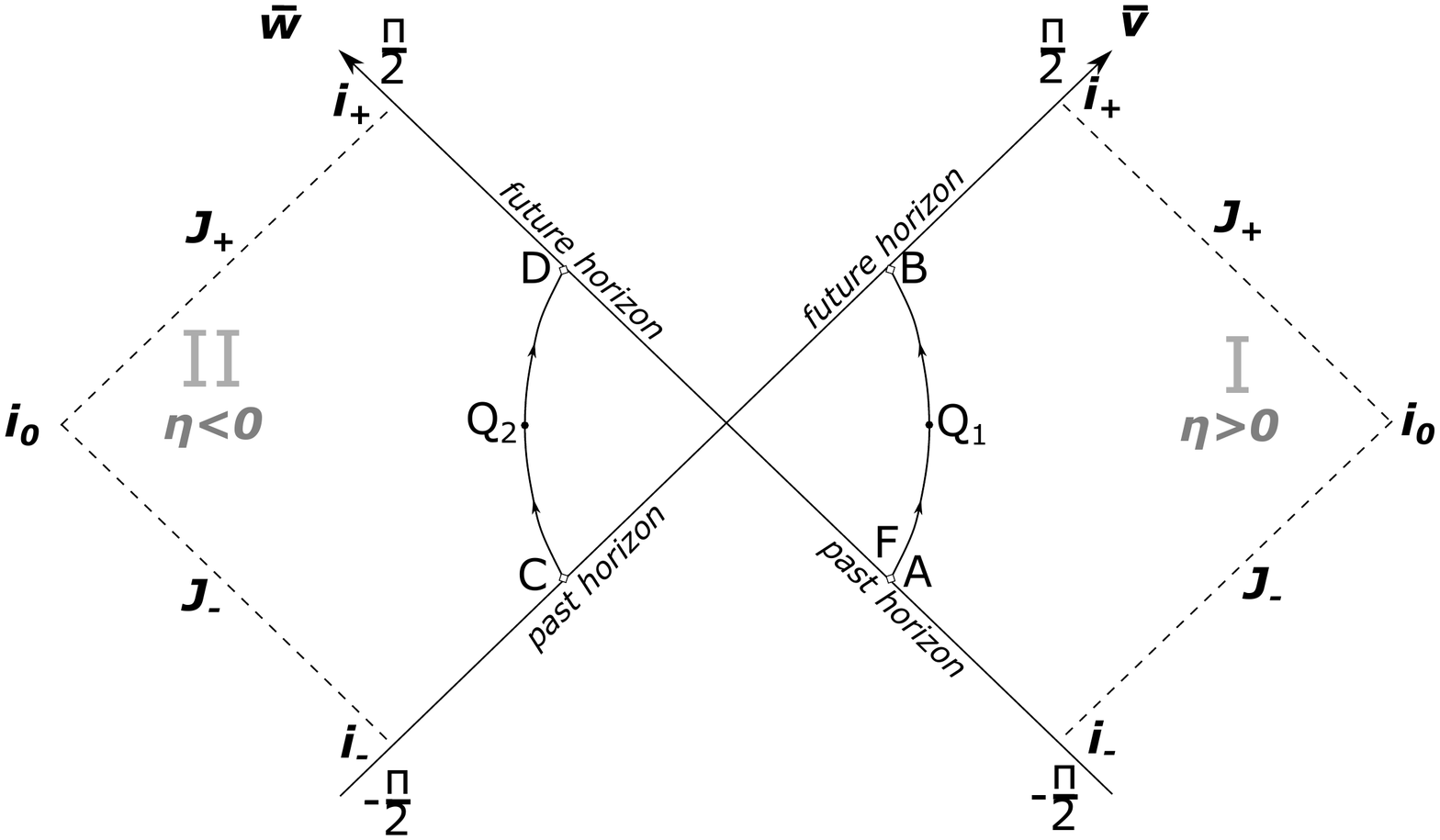}
\caption{Geodesic test-particle trajectory $A \to Q_1 \to B \sim C \to Q_2
\to D \sim F$. Eq.\rf{transcedent-eq} is the condition for 
$F=A$, \textsl{i.e.}, the geodesics is a CTC.}
\end{center}
\end{figure}

To describe the above geodesic curve $\bigl({\bar v}(\t),{\bar w}(\t)\bigr)$
on the Kruskal-Penrose coordinate chart (recall Eqs.\rf{Penrose-coord}) we
need the explicit expressions for the integrals \rf{t-plus-eta-star-2}
(regarded as functions of $x\equiv 2k_h \eta$):
\br
2k_h \bigl( t\pm\eta^{*}\bigr) (x) = \int dx \Bigl( 1+ \frac{1}{|x|}\Bigr)
\llb \eps \sqrt{\frac{1+|x|}{1+\bigl(1-\frac{m_0^2}{E^2}\bigr)|x|}} \pm 1 \rrb
\nonu \\
= f^{(\eps)}_{\pm} (x) + c^{\eps}_{\pm} \quad , \quad 
f^{(-1)}_{\pm} (x) = - f^{(1)}_{\mp} (x) \; .
\lab{t-plus-eta-star-3}
\er
Here:
\br
f^{(1)}_{\pm} (x) = \pm x - \frac{1}{b}\sqrt{(1+x)(1-bx)} + 
\log\bigl\lb 2+x-bx -2 \sqrt{(1+x)(1-bx)}\bigr\rb 
\nonu \\
+ \frac{1+3b}{2b^{3/2}} \arctan\Bigl(\frac{2bx+b-1}{b(1+x)(1-bx)}\Bigr)
+ \left(\begin{array}{c} 0 \\
 -2 \log x \end{array} \right) \phantom{aaaaaa} \; ,
\lab{f-def}
\er
where the short hand notation $b$ \rf{b-def} was used,
and $c^{\eps}_{\pm}$ are integration constants determined from the 
matching conditions at the points of return \rf{eta-stop} in regions $I$ 
and $II$: $x = \pm 2k_h \eta_{\rm stop} = \pm 1/b$ (using notation
\rf{b-def}).

Using Eqs.\rf{t-plus-eta-star-3}-\rf{f-def} the condition for existence of 
CTC -- coincidence on the past horizon of
region $I$ of the starting point of the particle geodesics $A$ with the 
endpoint of the same geodesics $F$ -- yields the following condition 
on the value of the parameter $b$ \rf{b-def}:
\be
\frac{1}{b} - \log(b+1) + \log 4 + \frac{1+3b}{2b^{3/2}}\Bigl\lb \frac{\pi}{2}
- \arctan\Bigl(\frac{b-1}{\sqrt{b}}\Bigr)\Bigr\rb = 0 
\lab{transcedent-eq}
\ee
with a solution $b \approx 5.5876$, \textsl{i.e.}
$m_0 \approx 2.5666\, E$.

Eq.\rf{transcedent-eq} implies for the intergration constants:
\be
f^{(1)}_{(-)}(0) + c^{(1)}_{(-)} = 2k_h \bigl( t-\eta^{*}\bigr)(0) = 0
\;\; ,\;\;
f^{(-1)}_{(+)}(0) + c^{(-1)}_{(+)}= 2k_h \bigl( t+\eta^{*}\bigr)(0)= 0 
\; ,
\lab{f-0-eqs}
\ee
where in the last equalities in \rf{f-0-eqs} the definition 
\rf{t-plus-eta-star-3} of $f^{(\pm 1)}_{(\mp)}(x)$ has been taken into 
account. 

Relations \rf{f-0-eqs} show that, for the CTC exhibited above, the points
$A,C$ and $B,D$ are located exactly at the middle of the past/future
horizons of regions $I$ and $II$:
\br
\bigl({\bar v}_A,{\bar w}_A\bigr)=\bigl(0,-\frac{\pi}{4}\bigr) \quad,\quad
\bigl({\bar v}_B,{\bar w}_B\bigr)=\bigl(\frac{\pi}{4},0 \bigr) \; ,
\nonu \\
\bigl({\bar v}_C,{\bar w}_C\bigr)=\bigl(-\frac{\pi}{4},0 \bigr) \quad,\quad
\bigl({\bar v}_A,{\bar w}_A\bigr)=\bigl(0,\frac{\pi}{4}\bigr) \; .
\lab{ABCD-locations}
\er

Existence of CTC's (also called ``time machines'') turns out to be quite 
typical phenomenon in wormhole physics (not necessarily of thin-shell
wormholes) \ct{NEC} (for a review, see \ct{CTC-lobo}). 
This is due to the violation of the null-energy conditions in
general relativity because of the presence of an ``exotic matter'' at the
``throat''. In the present
case of the original Einstein-Rosen bridge as a specific example of a
lightlike thin-shell wormhole, the violation of the null-energy conditions
is manifesting itself via the negativity of the dynamical lightlike membrane 
tension $T$ \rf{rel-param}, \textsl{i.e.}, the lightlike membrane
residing on the wormhole throat is an ``exotic'' lightlike thin-shell 
matter source.

\section{Conclusions}
\label{conclude}

We have discussed in some detail the basic properties of the mathematically
consistent formulation of the original ``Einstein-Rosen bridge'' proposed in
their classic 1935 paper. We have stressed a crucial feature (overlooked in
the 1935 paper) of the correctly formulated original Einstein-Rosen bridge 
-- it is {\em not} a solution of the vacuum Einstein equations but rather 
it the simplest example of a static spherically symmetric {\em traversable} 
lightlike ``thin-shell'' wormhole solution in general relativity. 
The consistency of the latter is guaranteed by the remarkable special 
properties of the world-volume dynamics of the lightlike membrane located 
at the wormhole ``throat'', which serves as an ``exotic'' thin-shell 
matter source of gravity.

\begin{itemize}
\item
We have described the Kruskal-Penrose diagram representation of the original
Einstein-Rosen bridge.
\item
The Kruskal-Penrose manifold of the original Einstein-Rosen bridge differs
significantly from the well-known Kruskal-Penrose extension of the standard
Schwarzschild black hole. Namely, Kruskal-Penrose coordinate chart of the 
original Einstein-Rosen bridge has only two regions corresponding to ``our''
(right) and the second (left) ``universes'' (two identical copies of the 
exterior Schwarzschild spacetime region) unlike the four regions in the
standard Kruskal-Penrose chart of the Schwarzschild black hole,
\textsl{i.e.}, now there are no black/white hole regions.
\item
There is a special pairwise identification between the future and past
horizons of the neighbouring Kruskal-Penrose regions.
\item
We have explicitly exhibited traversability of the original Einstein-Rosen
bridge w.r.t. ``proper-time'' of test-particles (travelling observers). In
particular, we have found for a special relation between the energy and the
mass of the test-particle a solution for the pertinent geodesics which is a
{\em closed timelike curve}. The latter is a typical feature in wormhole
physics with ``exotic'' matter sources.
\end{itemize}

\section*{Acknowledgements}
We gratefully acknowledge support of our collaboration through the academic 
exchange agreement between the Ben-Gurion University and the Bulgarian Academy 
of Sciences. S.P. and E.N. have received partial support from European COST 
actions MP-1210 and MP-1405, respectively. E.G. received partial support from 
COST Action CA-15117. E.N., S.P. amd M.S. gratefully acknowledge support from
Bulgarian National Science Fund Grant DFNI-T02/6. 


\end{document}